\UseRawInputEncoding
\documentclass[namedreferences]{solarphysics}
\usepackage[hyperref,optionalrh,solaromanenum]{spr-sola-addons} 
\usepackage{graphicx}                    
\usepackage{amssymb}                    
\usepackage{color}                       
\usepackage{breakurl}                    

\begin{document}


\begin{opening}

\title{Comparison of Geomagnetic Indices During Even and Odd Solar Cycles SC17\,--\,SC24: Signatures of Gnevyshev Gap in Geomagnetic Activity
}

%
\author[addressref={aff1},corref,email={jojuta@gmail.com}]{\inits{J.T.}\fnm{Jouni}~\lnm{Takalo}}

\institute{$^{1}$ Space Physics and Astronomy Research Unit, University of Oulu, POB 3000, FIN-90014, Oulu, Finland
}
%
\runningauthor{J.J. Takalo}
\runningtitle{Geomagnetic Disturbances During Solar Cycles 17\,--\,23}



\begin{abstract}
We show that the time series of sunspot-group areas has a gap, the so-called Gnevyshev gap (GG), between ascending and descending phases of the cycle and especially so for the even-numbered cycles. For the odd cycles this gap is less obvious, and is only a small decline after the maximum of the cycle. We resample the cycles to have the same length of 3945 days (about 10.8 years), and show that the decline is between 1445\,--\,1567 days after the start of the cycle for the even cycles, and extending sometimes until 1725 days from the start of the cycle. For the odd cycles the gap is a little earlier, 1332\,--\,1445 days after the start of the cycles with no extension.
We analyze geomagnetic disturbances for Solar Cycles 17\,--\,24 using the Dst-index, the related Dxt- and Dcx-indices, and the Ap-index. In all of these time series there is a decline at the time, or somewhat after, the GG in the solar indices, and it is as deepest between 1567\,--\,1725 days for the even cycles and between 1445\,--\,1567 days for the odd cycles. The averages of these indices for even cycles in the interval 1445\,--\,1725 are 46\,\%, 46\,\%, 18\,\%, and 29\,\% smaller compared to surrounding intervals of similar length for Dst, Dxt, Dcx, and Ap, respectively. For odd cycles the averages of the Dst- and Dxt-indices between 1322\,--\,1567 days are 31\,\% and 12\,\% smaller than the surrounding intervals, but not smaller for the Dcx-index and only 4\,\% smaller for the Ap-index. The declines are significant at the 99\,\% level for both even and odd cycles of the Dst-index and for Dxt-, Dcx- and Ap-indices for even cycles. For odd cycles of the Dxt-index the significance is 95\,\%, but the decline is insignificant for odd cycles of the Dcx- and Ap-indices. 
\end{abstract}

%

\end{opening}


\section{Introduction}
The solar cycle has basically three phases: an ascending phase, a descending phase, and between them a so-called Gnevyshev gap (GG: \cite{Antalova_1965, Gnevyshev_1967, Gnevyshev_1977, Feminella_1997, Storini_2003, Ahluwalia_2004, Kane_2005, Bazilevskaya_2006, Kane_2008a, Norton_2010, Du_2015, Takalo_2018}), which is a kind of separatrix between the first two (main) phases.  The time of the the Gnevyshev gap is 45\,--\,55 months after the start of the nominal cycle, that is, approximately 33\,--\,42\,\% into the cycle after its start \citep{Takalo_2018}.

\cite{Storini_2003} gave a review of the effects of the GG in different space-weather parameters. There are many studies about the overall correlation of the solar and interplanetary activity measure with different aspects of geomagnetic activity. Some of the recent analyzes are mentioned in the following discussion.

\cite{Feminella_1997} examined the long-term behavior of several solar-activity parameters (spots, flares, radio-,
and X-ray fluxes), confirmed the double peaks, and found that the peaks are more distinct with a clear gap in between when intense and/or long-lasting events are considered. On the other hand, low-energy and short-duration events tend to follow a single-peaked 11-year cycle. The double peaks are reported to be detected in all the solar atmospheric layers (photosphere, chromosphere, and corona) up to interplanetary space and are linked with the heliomagnetic cycle. Double peaks have also been detected in geomagnetic activity \citep{Gonzalez_1990, Kane_1997, Echer_2004}. Interestingly \cite{Ahluwalia_2000} found, while studying the connection between IMF parameters, that solar polar-field reversals may be responsible for the existence of the Gnevyshev gap in Ap-data. A function proportional to the IMF magnetic field intensity multiplied with square of solar-wind velocity [$BV^{2}$] describes the temporal variations of Ap quite well. \cite{Kane_2002} showed that the evolution of various solar indices around sunspot maximum occurs almost simultaneously (within a month or two) and the gap between the two peaks is shown to be sharper for more energetic events. \cite{Takalo_2020} and \cite{Takalo_2020_2} have also shown that the larger the sunspot groups the more clearly their overall distribution during the solar cycle is double-peaked.

\cite{Verbanac_2011} analyzed the correlation (during 1960\,--\,2001) of several solar-activity indices: sunspot number, group sunspot number, cumulative sunspot area, solar radio flux $F_{10.7}$, and interplanetary magnetic field strength (IMF) with geomagnetic activity indices: Ap, Dst, and Dcx, and the regional geomagnetic index RES, specifically estimated for the European region. The geomagnetic indices Ap, Dcx, Dst, and RES are all well correlated with the considered solar indices. Interestingly, RES showed the most prominent correlation with solar indices. All of the geomagnetic indices were found to have the best correlation with the IMF. The highest (anti)correlation was found for RES vs. IMF (correlation coefficient, R = \textrm{-}0.77) for one-year time lag. Ap is also delayed for one year after IMF, while Dcx and Dst are synchronized with IMF. A two-year lag is found between all solar indices (except IMF) and Ap.
                                                                                  
\cite{Kilcik_2017} studied different data sets from January 1996 to March 2014 and found that the best correlation between the sunspot counts and the Ap- index is obtained for the large-group time series, but they noted that for the highest correlation the Ap-index is delayed 13 months compared to all kinds of sunspot-count series and International sunspot number (ISN) data. They also found that the best correlation between the sunspot counts and the Dst-index is obtained for the large sunspot-group time series. The Dst-index delays with respect to the large-group by about two months. Furthermore, the correlation coefficients between the geomagnetic indices (Ap, Dst) and X-ray solar-flare index are a little higher than the correlation coefficients between these geomagnetic indices and ISN.

\cite{Le_2013} examined occurrence of intense storms at the level of \textrm{-}200\,nT $<$ Dst $\leq$ \textrm{-}100\,nT, great storms at \textrm{-}300\,nT $<$ Dst $\leq$ \textrm{-}200\,nT, and super storms at Dst $\leq$ \textrm{-}300\,nT during the period 1957\,--\,2006, based on Dst-indices and smoothed monthly sunspot numbers. Their statistics showed that 76.9\,\% of the intense storms, 79.6\,\% of the great storms and 90.9\,\% of the super storms occurred during the two years before a solar cycle reached its peak, or in the three years after it. The histograms of yearly number of intense geomagnetic storms have a dual-peak distribution in Solar Cycles 21 and 22, and three peaks in Solar Cycles 20 and 23. The first peak in yearly number of intense geomagnetic storms occurred one year before the solar cycle maximum in Solar Cycles 20, 21, and 23, but only in Solar Cycle 22 did the first peak occur at the solar maximum. The second peaks for Solar Cycles 20\,--\,23 all occurred two years after the solar maximum. The gap between the the first and second peaks for intense geomagnetic storms is from one to three years. Note, however, that their histograms are very sparse, and the width of each bin is one year.

Less attention has been paid to the effects of GG on the geomagnetic activity at the Earth. \cite{Richardson_2012_1} and \cite{Richardson_2013} have, however, performed studies about the drop of geomagnetic disturbance during the maximum of the solar cycle. They found, especially, a tendency for the Kp-index storm rate to fall during the fourth year of the cycle. They also associated several occasions of decline in geomagnetic activity aa-index around solar maximum with the Gnevyshev gap during Solar Cycles 20\,--\,24\ \citep{Richardson_2012_2}

In this study we analyze the strength of geomagnetic activity using daily time series of Dst for years 1957\,--\,2009, extended Dst \citep{Karinen_2005}, Dxt for 1933\,--\,2009, corrected and extended Dst \citep{Karinen_2006}, Dcx for 1933\,--\,2009, and Ap-index for years 1933\,--\,2020. We first find the interval of the Gnevyshev gap in sunspot-number and sunspot-area data, and then we analyze the statistical significance for the decline during or after the same interval in geomagnetic indices. (Note especially that we do not study cross-correlation of the solar and  geomagnetic indices themselves.) This article is organized as follows: Section 2 presents the data used in this article. In Section 3 we study sunspot-number and sunspot-area data for Solar Cycles (SCs) 17\,--\,23, and also for the longer period of SCs 12\,--\,23 to get better statistics. In Section 4 we analyze the Dcx, Dxt, and Dst during Solar Cycles 17\,--\,23, and in Section 5 the Ap-index during Solar Cycles 17\,--\,24. In Section 6 we discuss the signatures of GG in the near-Earth interplanetary magnetic field and solar-wind velocity using OMNI2 daily data. In Section 7 we consider the sequence of group areas, IMF $BV^{2}$, and Ap-index for SC20\,--\,SC24 and give our conclusions in Section 8.

\section{Data}

\subsection{Solar Group Area and OMNI2 Data}

In the analysis of sunspot groups for Solar Cycles 12\,--\,24 we use the recently published catalogue of sunspot groups by \cite{Mandal_2020}. The main data for this data set are the Royal Greenwich observatory (RGO) measurements between 1874\,--\,1976. The years after 1976 until October 2019 were constructed from the data sets from the Kislovodsk, Pulkovo, and Debrecen observatories. This data set contains the daily appearance of the sunspot groups. Thus the lifetime of each group also affects to the number of the groups. The minima and length of the sunspot cycles used in this study for solar and also geomagnetic indices are listed in Table 1. (We are aware that geomagnetic indices lag somewhat the solar indices, but for this study using the same cycles is appropriate.)

\begin{table}
\small
\caption{Sunspot-Cycle lengths and dates [fractional years, and year and month] of (starting) sunspot minima for Solar Cycles 12\,--\,23 (except the end of data for Solar Cycle 24). \citep{NGDC_2013}.}
\begin{tabular}{ c  c  l  c }
  Sunspot cycle    &Fractional    &Year and month     &Cycle length  \\
      number    &year of minimum   & of minimum     &    [years] \\
        \hline   
12    & 1879.0  &1878 December  & 10.6  \\
13    & 1889.6  &1889 August  & 12.1  \\
14    & 1901.7  &1901 September  & 11.8  \\
15    & 1913.5  &1913 July  & 10.1  \\
16    & 1923.6  &1923 August & 10.1  \\
17    & 1933.7  &1933 September  & 10.4  \\
18    & 1944.1  &1944 February  & 10.2  \\
19    & 1954.3  &1954 April  & 10.5  \\
20    & 1964.8  &1964 October  & 11.7  \\
21    & 1976.5  &1976 June & 10.2  \\
22    & 1986.7  &1986 September  & 10.1  \\
23    & 1996.8  &1996 October  & 12.2  \\
24    & 2009.0  &2008 December  & 10.9  \\ 
25    & 2019.8  &2019 October 
\end{tabular}
\end{table}

The OMNI2 data set contains the hourly mean values of the interplanetary magnetic field (IMF) and solar-wind plasma parameters 
measured by various spacecraft near the Earth's orbit, as well as geomagnetic and solar-activity indices, and energetic-proton fluxes. In our analysis we use daily averages of the absolute value of the magnetic-field intensity \textit{B} and solar-wind bulk velocity for the years 1964\,--\,2020 (Solar Cycles 20\,--\,24).

\subsection{Geomagnetic Indices}

The Dst (storm-time disturbance) index has been calculated at the World Data Center C2 at Kyoto, Japan, since the International Geophysical Year 1957, using data from four observatories at low latitudes (Hermanus, Honolulu, Kakioka, and San Juan).

The Dxt-index is an extension of Dst using data from the original set of four low-latitude stations for 1941\,--\,1956, and using the nearby Cape Town station as a predecessor of the Hermanus station for 1932\,--\,1940 \citep{Karinen_2005}. Despite some open questions related to the composition of the original Dst-index, the reconstructed index is quite similar to the original one during the overlapping time interval (1957\,--\,2002). However, the reconstructed Dxt-index corrects for some known errors in the original Dst-index, such as the erroneously large daily UT variation in 1971 \citep{Takalo_2001, Karinen_2002}.

Dcx-index is an extended and corrected version of the Dst-index such that it also corrects the Dst-(and Dxt-)index for the excessive, seasonal varying quiet-time level, the so-called ``non-storm component", which is unrelated to magnetic storms. Thus the Dcx-index has smaller seasonal variation than the other indices \citep{Mursula_2005, Karinen_2006}.

Planetary geomagnetic activity index [Ap] is derived from a range of geomagnetic field variations over a period of three hours from measurements provided by 13 geomagnetic observatories between \textrm{-}44$^{\circ}$ and 60$^{\circ}$ southern and northern geomagnetic latitude. We use here daily averages of the Ap-index, which is a measure of the daily level of the mid-latitude magnetic activity. The Ap-index exists for Solar Cycles 17\,--\,23, and nowadays also, at least as preliminary values, for Solar Cycle 24.



\section{Sunspot Area Data for Solar Cycles 12\,--\,23}

\begin{figure}
	\centering
	\includegraphics[width=1.0\textwidth]{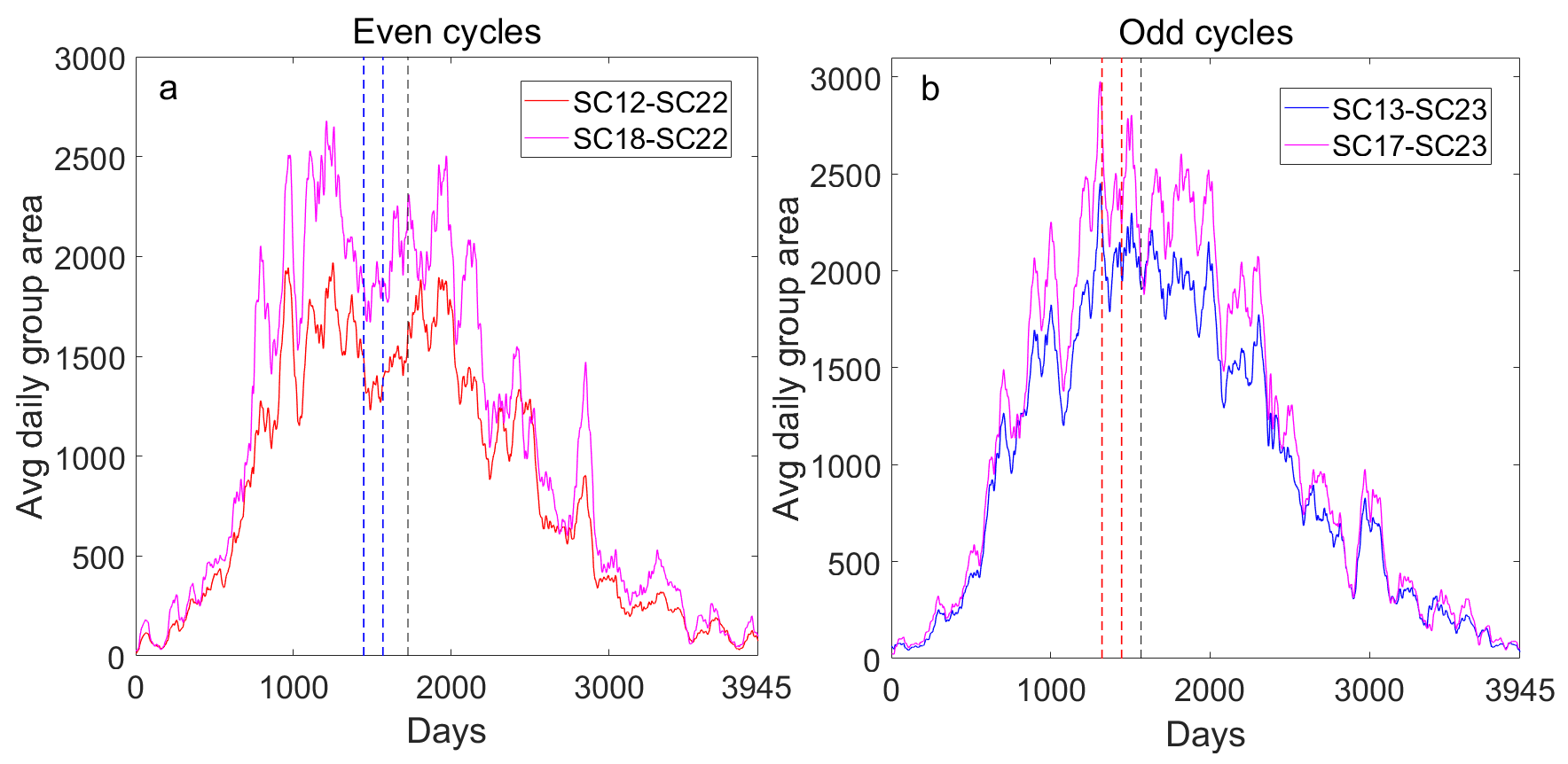}
		\caption{Average daily sunspot-group area for a) the even Solar Cycles 12\,--\,22 and 18\,--\,22. b) for the odd Solar Cycles 13\,--\,23 and 17\,--\,23.}
		\label{fig:Sunspot_area}
\end{figure}

\begin{figure}
	\centering
	\includegraphics[width=1.0\textwidth]{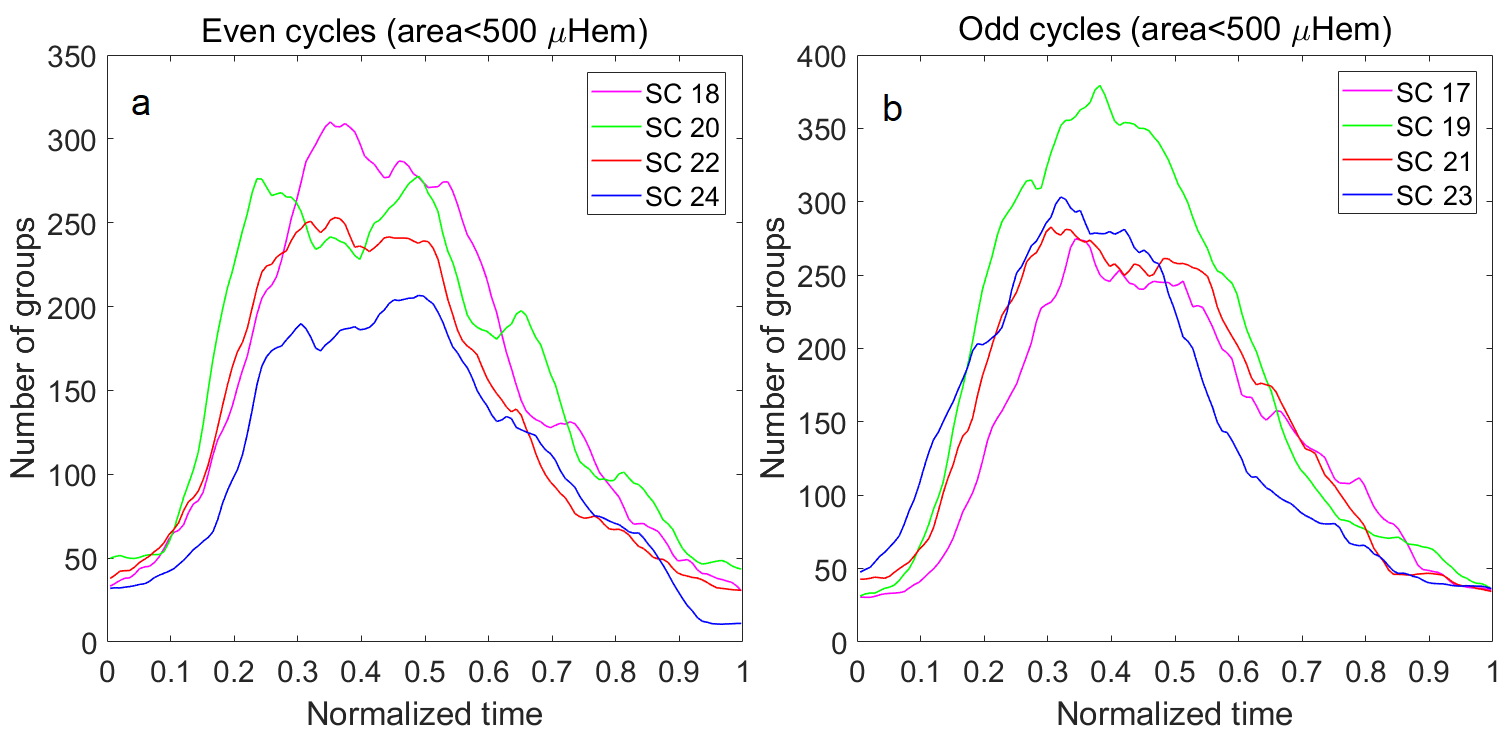}
		\caption{Number of sunspot groups with area$<$500$\mu$Hem a) for the even Solar Cycles 18\,--\,24. b) for the odd Solar Cycles 17\,--\,23.}
		\label{fig:areas_smaller_500}
\end{figure}

Figures \ref{fig:Sunspot_area}a and b show the total sunspot areas during Solar Cycles 12\,--\,23 for the even and odd cycles, respectively. The area data are smoothed using trapezoidal smoothing over 61 points (rectangular moving-average smoothing with window end points having only half of the weight of the inner points). Also, all of the other data are smoothed similarly if not otherwise mentioned. It is evident that the average even cycle is double-peaked such that there is a clear decline between the peaks at about days 1445\,--\,1567 (marked with blue dashed lines) and a weaker decline for the prolonged interval 1567\,--\,1725 days (end point with fainter black dashed line). This is especially true for the longer period of Solar Cycles 12\,--\,22. We believe that the gap is related to the Gnevyshev gap (GG). In contrast, the average odd cycle has only one maximum (peak), after which (at about day 1322) the sunspot index gradually decreases, having a sudden short gap between days 1322\,--\,1445 (the red dashed lines). The prolonged interval marked for odd cycles is days 1445\,--\,1567. We could, however, say that the GG is only the start of the descending phase for the odd cycles. One reason why the GG is more intense in the even cycles is that the even cycles reach their maximum on average at different times in the southern and northern solar hemispheres, and although GG exists in both hemispheres, there is a longer and deeper GG for the even cycles \citep{Takalo_2020, Takalo_2020_2}. Note also a (quasi)periodic structure in the average daily area time series. The mean period of this fluctuation is about 150 days, which has been found to be the period of various activities of the Sun \citep{Rieger_1984, Lou_2000, Richardson_2005}. 
The decline near the cycle maximum is usually due to larger groups, which have this double-peak pattern during the solar cycle. Figure \ref{fig:areas_smaller_500}a and b shows the number of smaller groups (area$<$500 $\mu$Hem, millionths of hemisphere) for the even and odd cycles of SC17\,--\,SC24 as a function of normalized time (all cycles are normalized such that $T_{\texttt{norm}}=\left(T_{i}-T_{\texttt{min}}\right)/L$, where $T_{i}$ is the original time of each group, $T_{\texttt{min}}$ is the time for the leading minimum of the cycle, and $L$ is the length of the cycle). The data are calculated from monthly averages and smoothed using 13-month trapezoidal method. It is seen that all odd cycles have only one clear maximum, and only Solar Cycle 20 from the even cycles has distinct two-peak structure. This result is in line with the earlier studies mentioned in the Introduction.

\section{Dst, Dxt, and Dcx During Cycles 17\,--\,23}

\begin{figure}
	\centering
	\includegraphics[width=1.0\textwidth]{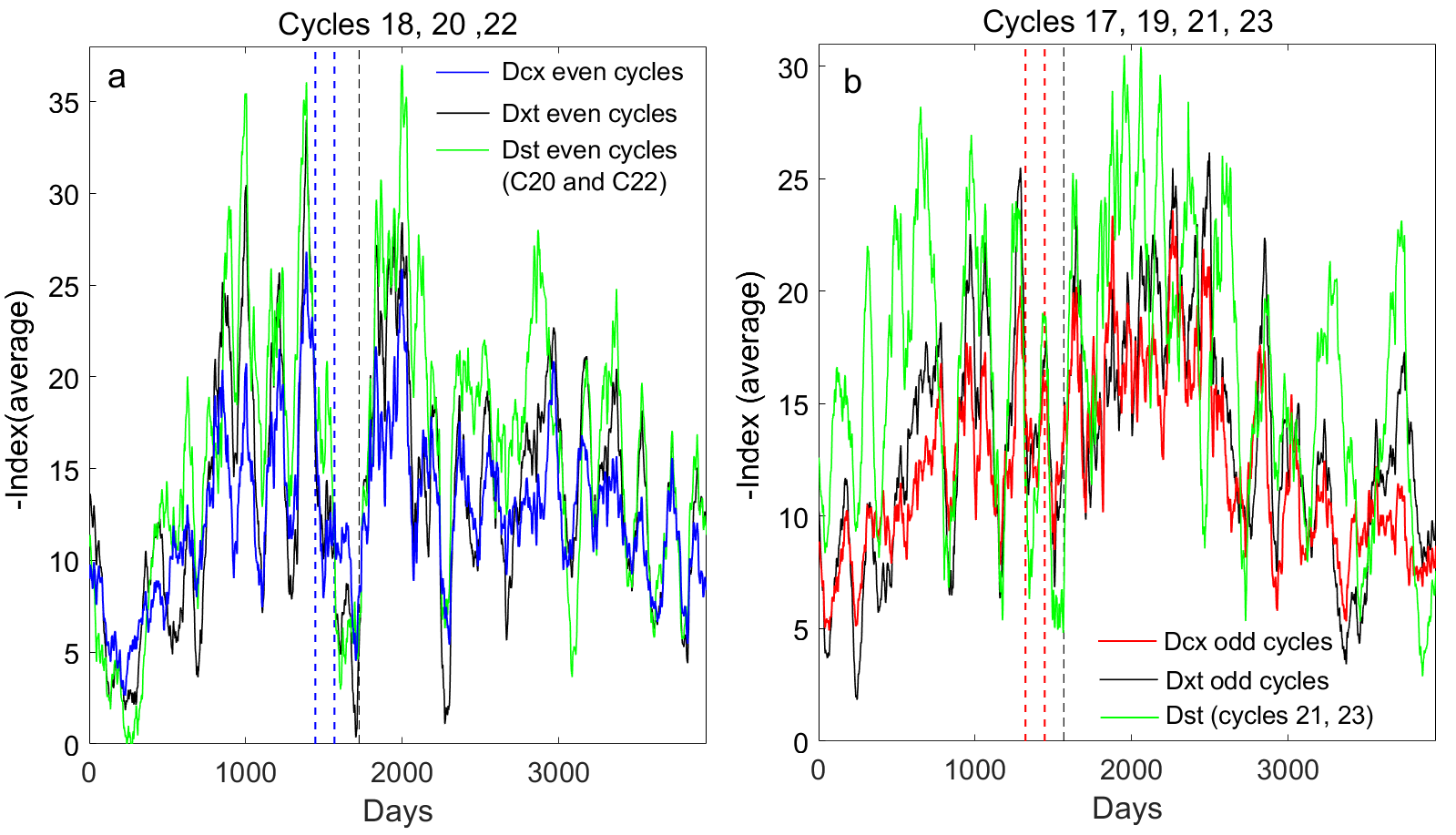}
		\caption{a) The average value (opposite sign) of the smoothed Dst, Dxt, and Dcx indices of the even Solar Cycles 18, 20, and 22. b) Same for the odd Solar Cycles 17, 19, 21, and 23.} 
		\label{fig:Dcx_Dxt_Dst_averages}
\end{figure}

Positive values of Dst are mainly due to the compression of the dayside magnetosphere in the initial phase of a storm, while negative values are due to magnetic reconnection and the formation of the storm-related currents, in particular the ring current. Therefore positive and negative values of the Dst-index arise from different physical processes. That is why we use in the next analysis only negative values of the Dst and related indices (The positive values are treated as zero. We verified that this does not affect the results in this study. If the original time series is used, it is separately mentioned.)
We then resampled Dcx-, Dxt-, and Dst- indices for each Solar Cycle 17\,--\,23 (Dst only for Solar Cycles 20\,--\,23) such that they all had same length: 3945 time steps (days). This is about 10.8 years, which is the average of cycles since the beginning of the 20th century. Figures \ref{fig:Dcx_Dxt_Dst_averages}a and \ref{fig:Dcx_Dxt_Dst_averages}b show the averages of the absolute values of the Dcx, Dxt, and Dst for even and odd cycles, respectively. Here the indices are smoothed using trapezoidal smoothing over 61 points. Note that the Dst-index exists only since 1957, and that is why we can show only total Solar Cycles 20\,--\,23 for Dst.
The vertical dotted lines are at 1445, 1567, and 1725 days, and at 1322, 1445, and 1567 for the average even and odd cycle, respectively. It is evident that for the even cycles the indices have a decline in the whole interval 1445\,--\,1725 days, but are deepest at the end of the nine-month period. This interval starts about 37\,\% after the start of the average cycle. It is also evident that the deepest decline lags some months the Gnenyshev gap (GG) in the sunspots, which is located between the days 1445 and 1567 (see Figure \ref{fig:Sunspot_area}). For odd cycles there is a decline but somewhat earlier, i.e. between 1322\,--\,1567, and slightly deeper in the latter half (1445\,--\,1567) of the marked interval. However, the decline is not so clear for the odd cycles as for the even cycles.
 
We calculated, using two-sided T-test for unequal means (see the T-test method in \cite{Takalo_2020_1} and \cite{Takalo_2020}), that the significance of the decline (average Dst is \textrm{-}12.7) for the raw (unsmoothed) Dst-index is 99\,\% for even cycles (SC20 and SC22) with p\,$<\,10^{-15}$ compared to nine months earlier (average Dst \textrm{-}23.8) and nine months after the marked interval (average Dst \textrm{-}23.6). For odd cycles (SC21 and SC23) the decline between 1322\,--\,1527 is \textrm{-}11.9 on the average with p\,=\,$3\times10^{-5}$ compared to eight months earlier (average Dst \textrm{-}17.3) and p\,=\,$9\times10^{-9}$ compared to eight months after the marked interval (average Dst is \textrm{-}19.5). 
For Dxt-index the corresponding 99\,\% significance exists with p-value $<\,10^{-18}$ for even Solar Cycles 18, 20, and 22 between 1445\,--\,1725 (average Dxt is \textrm{-}11.2) compared to the interval before (average Dxt \textrm{-}20.9) and after (average Dxt \textrm{-}22.2). However, for odd Solar Cycles 17, 19, 21, and 23, the significance is only at the 95\,\% level with p-value 0.017 (the average Dxt-values are \textrm{-}17.2, \textrm{-}15.1, and \textrm{-}17.4 for before, during, and after the interval 1322\,--\,1567, respectively).
This is probably because the odd Solar Cycles 21 and 23 have clearer double peaks than the earlier odd cycles, and consequently the decline is pronounced during these cycles (note that Dst-index exists only during odd Solar Cycles 21 and 23).
For the raw Dcx-index, the two-sided T-test shows 99\,\% significance with p\,$<\,10^{-9}$ for the even cycles between 1445\,--\,1725. The values of average Dcx are \textrm{-}19.3, \textrm{-}11.9 and \textrm{-}18.4 before, during, and after the interval, respectively.  For the odd cycles, the decline in the interval 1322\,--\,1567 is insignificant with the average values being \textrm{-}14.6, \textrm{-}14.8, and \textrm{-}16.8 for Dcx before, during, and after the interval 1322\,--\,1567, respectively. It seems that the equinox correction changes the profile during this interval for the Dcx. However, for Dcx-index$<$\textrm{-}50\,nT there is a significant decline compared to the interval after the marked period with p\,=\,0.0093.

\begin{figure}
	\centering
	\includegraphics[width=1.0\textwidth]{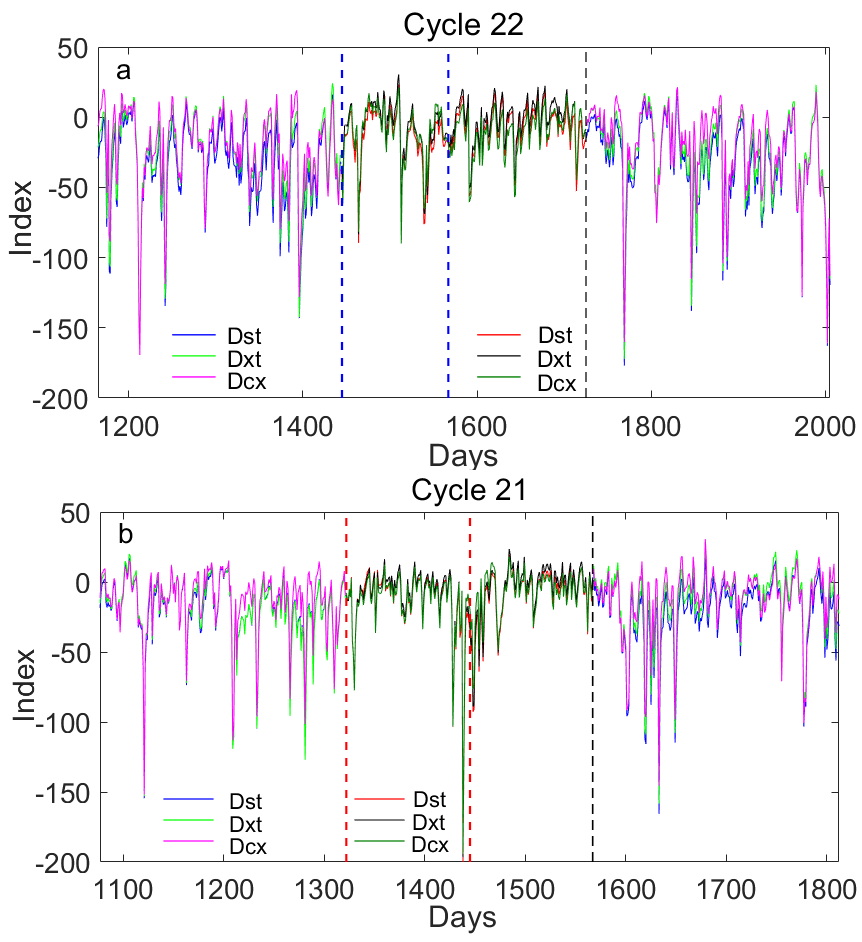}
		\caption{a) Part of the resampled unsmoothed (cycle length 3945 days) time series for Solar Cycle 22. b) Same, but for Solar Cycle 21.}
		\label{fig:Dcx_Dxt_Dst_indices}
\end{figure}

Figure \ref{fig:Dcx_Dxt_Dst_indices}a and b depict an example of what the decline looks like in the original indices for Solar Cycle 22 and Solar Cycle 21, respectively. Here we use the original unsmoothed indices with positive values included. Note that the time axis is similar to Figure \ref{fig:Dcx_Dxt_Dst_averages}, but only the interval 1165\,--\,2005 for even cycle and 1077\,--\,1812 for odd cycle are shown. Note also that the vertical dashed lines show the same intervals as in the earlier figures for the sum of the cycles. It is evident that the indices have smaller values between the dashed lines and especially in the latter interval of those lines. For the odd Solar Cycle 21 there is a strong magnetic storm just around the line at 1438, which lasts in total about 20 days and is also seen in the Figure \ref{fig:Dcx_Dxt_Dst_averages}b. This storm notably changes the behavior of the average curve at that site. Otherwise, the intervals between the dashed line are very quiet considering that it was time of the maximum of Solar Cycle 21.

\begin{figure}
	\centering
	\includegraphics[width=1.0\textwidth]{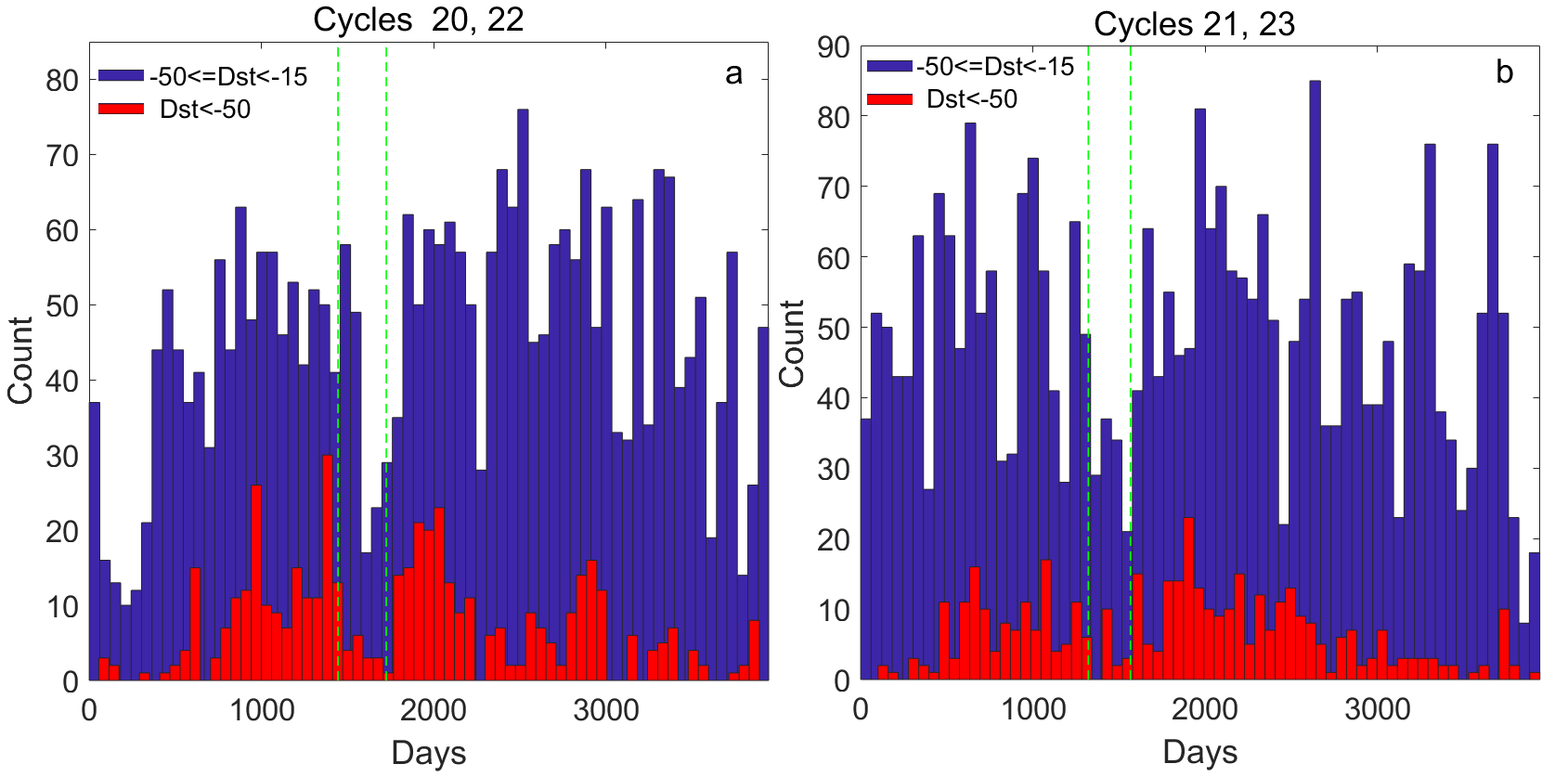}
		\caption{a) Histograms for the even Solar Cycles 20 and 22 of the Dst-index. b) Histograms for the odd Solar Cycles 21 and 23 of the Dst-index.}
		\label{fig:Dst_histogram}
\end{figure}

\begin{figure}
	\centering
	\includegraphics[width=1.0\textwidth]{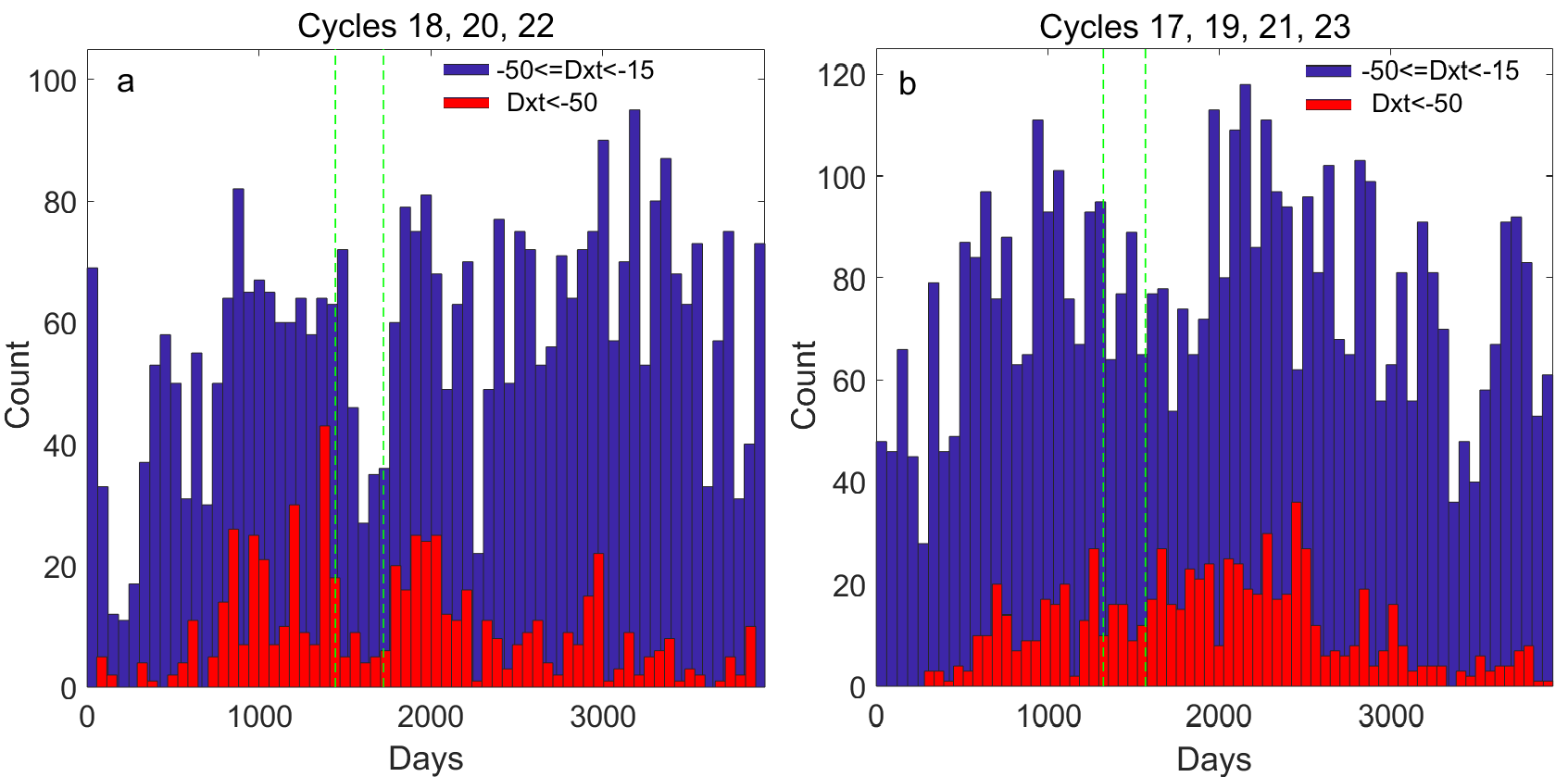}
		\caption{a) Histograms for the even Solar Cycles 18, 20, and 22 of the Dxt-index. b) Histograms for the odd Solar Cycles 17, 19, 21, and 23 of the Dxt-index.}
		\label{fig:Dxt_histogram}
\end{figure}

\begin{figure}
	\centering
	\includegraphics[width=1.0\textwidth]{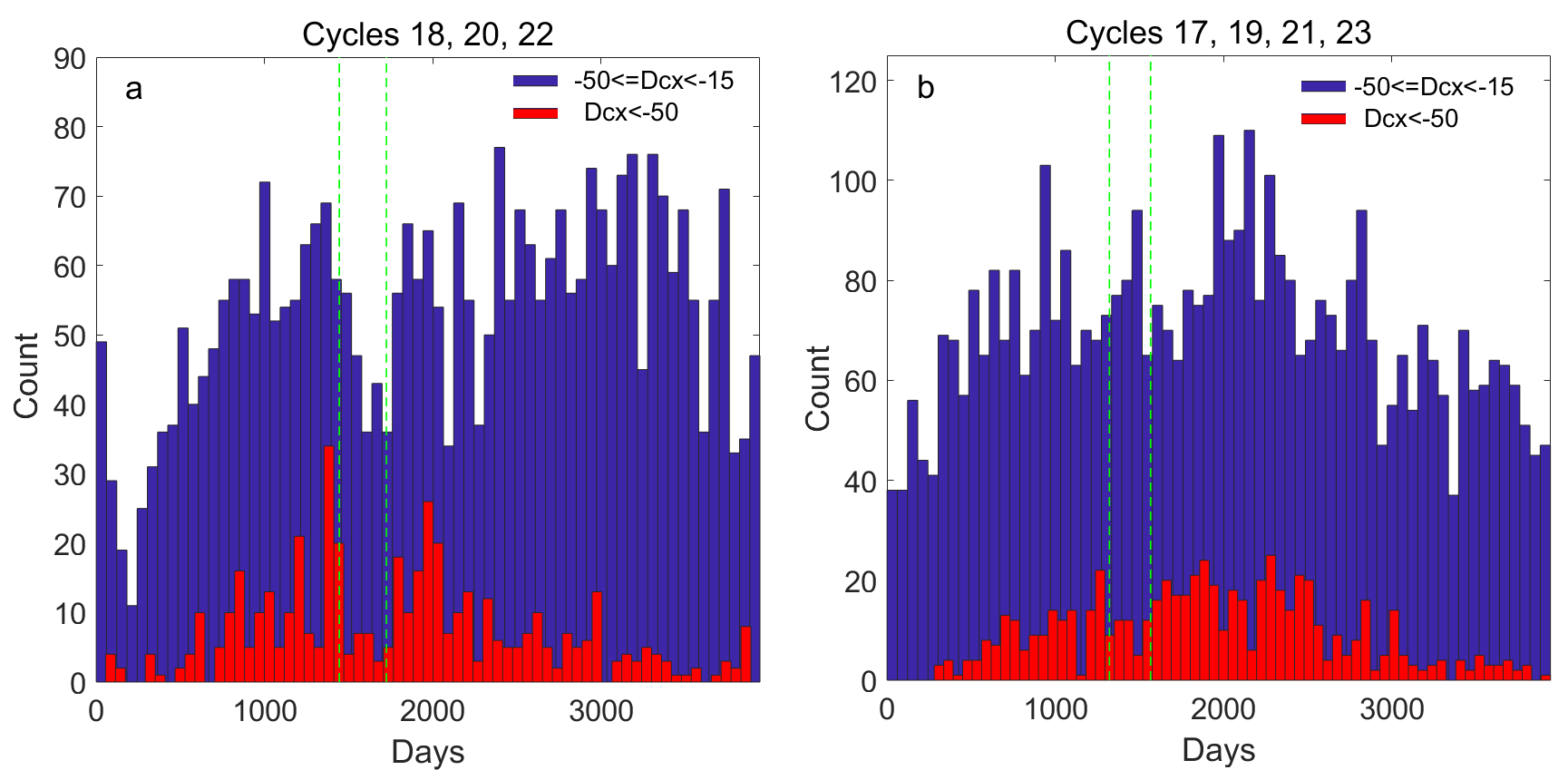}
		\caption{Histograms for the even Solar Cycles 18, 20, and 22 of the Dcx-index. b) Histograms for the odd Solar Cycles 17, 19, 21, and 23 of the Dcx-index.}
		\label{fig:Dcx_histogram}
\end{figure}

We have also divided the disturbed-time indices to two categories: strong, index$\leq$\textrm{-}50\,nT, and medium, \textrm{-}50\,nT$<$index$\leq$-15\,nT. (The category strong is defined as quite low level to get better statistics for our research. Note, however, that \textrm{-}50\,nT is about the mean plus twice the standard deviation for the original indices and \textrm{-}15\,nT is near the mean for the original Dst). Figures \ref{fig:Dst_histogram}a and b show the histograms of the two categories: index$<$\textrm{-}50\,nT and \textrm{-}50\,nT$\leq$index$<$-15\,nT for the even and odd Solar Cycles 20\,--\,23, respectively (Note that one bin is about two months). The vertical dotted lines are at 1445 and 1725 days (nine months) for even cycles and 1322 and 1567 days (eight months) for odd cycles. In both the even and odd cycles there is a decline between these lines, but it is emphasized in the latter part of this interval, and it is clearer for the even cycles than the odd cycles. Note that the minimum of the distribution lags by about four months in the medium-sized index category that of the GG decline in the sunspot indices.
Figures \ref{fig:Dxt_histogram}a and b show the histograms of the same categories for the Dxt-index. (Note that here we have twice as many cycles and the histograms show the total number of counts in each category.) For the odd cycles there is no decline either in index$\leq$\textrm{-}50\,nT or \textrm{-}50\,nT$\leq$index$<$\textrm{-}15\,nT category. This seems, however, to be due to Solar Cycles 17 and 19, which fill the gaps seen in odd Dst cycles (SC21 and SC23). For even cycles the Figures \ref{fig:Dst_histogram}a and \ref{fig:Dxt_histogram}a do not differ very much from each other. 
Figures \ref{fig:Dcx_histogram}a and b show the histograms of the same categories for the Dcx-index. The histograms in both categories of disturbances are very similar to those for the Dxt-index.

\section{Ap Index During Solar Cycles 17\,--\,24}

\begin{figure}
	\centering
	\includegraphics[width=0.9\textwidth]{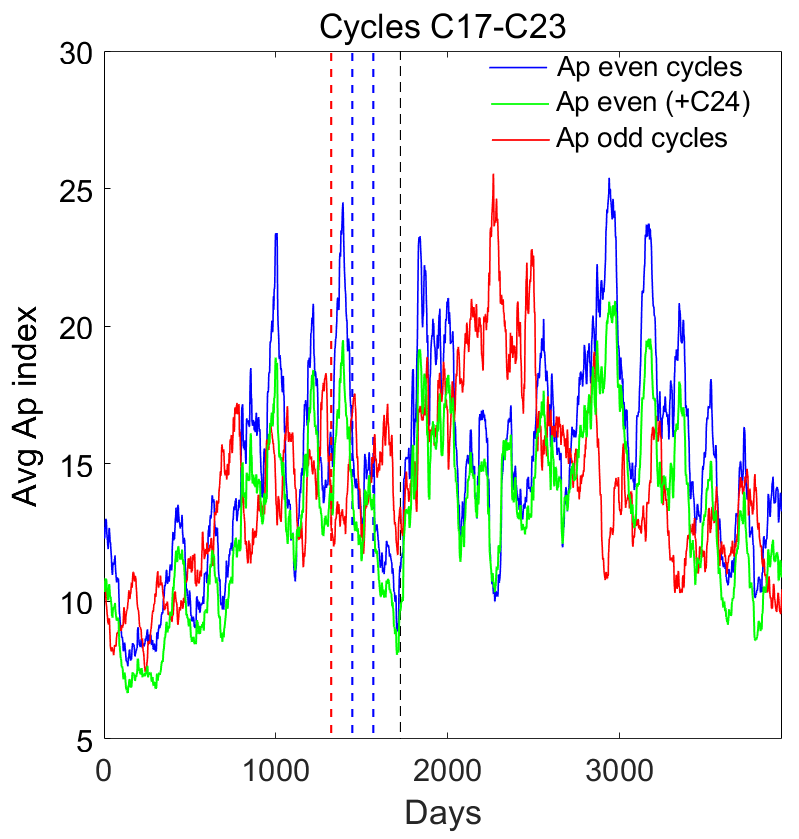}
		\caption{The average values of the smoothed Ap-index for the even Solar Cycles 18, 20, 22, and 24, and for the odd Solar Cycles 17, 19, 21, and 23}
		\label{fig:Ap_averages}
\end{figure}

Figure \ref{fig:Ap_averages} shows the resampled (3945 days) averages of the smoothed Ap-index for the even and odd cycles for Solar Cycles 17\,--\,24. (Vertical lines are again the same as in the earlier figures.) There is a decline in the even cycles, but mainly during the last interval between days 1567\,--\,1725. While the decline in solar sunspot-number and sunspot-group area data is between 1445\,--\,1567, this means that the decline in Ap is on average about four months later than in the Sun. The two-sided T-test shows that the whole interval 1445\,--\,1725 (nine months) has a lower mean value of Ap-index with p\,$<\,10^{-10}$ compared to both the nine month before and nine months after the interval (as calculated from the raw unsmoothed data). The mean value of Ap for the interval 1445\,--\,1725 is 12.7, while for nine months earlier (later) Ap is 18.0 (18.4). In contrast to this, the difference in odd cycles between 1322\,--\,1567 and eight months earlier or after the interval is not statistically significant. The value of Ap for this interval of odd cycles is 13.9, while for eight months earlier (later) it is 14.7 (14.5).

\begin{figure}
	\centering
	\includegraphics[width=1.0\textwidth]{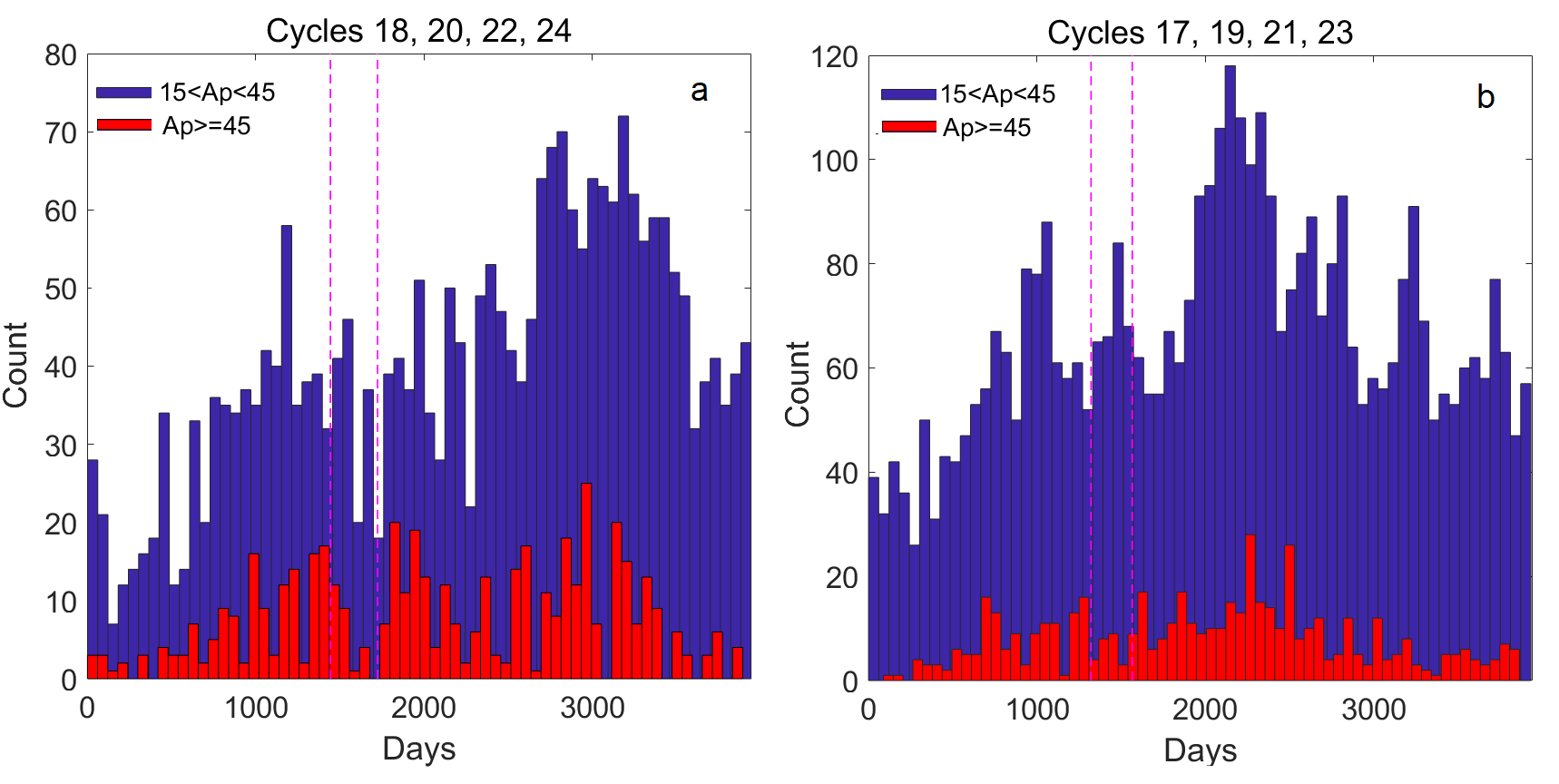}
		\caption{Histograms for the even Solar Cycles 18, 20, 22, and 24 of the Ap-index. b) Histograms for the odd Solar Cycles 17, 19, 21, and 23 of the Ap-index.}
		\label{fig:Ap_histogram}
\end{figure}

We have also divided the Ap-index in two categories: strong, index$>$45\,nT (mean plus twice standard deviation), and medium, 15\,nT$<$index$\leq$45\,nT.  Figures \ref{fig:Ap_histogram}a and b show the histograms of the two categories for the even and odd Solar Cycles 17\,--\,24, respectively. (Note that we have included Solar Cycle 24 in Figure \ref{fig:Ap_histogram}a). For the even cycles there seems to be a decline about four months after the GG interval in sunspots, but not a significant decline for odd cycles. Note that the profile of Ap-index greater than 15\,nT resembles very much that of aa-index \citep{Cliver_1996}, i.e. for the even cycles the distribution is highest in the declining phase of the sunspot cycle, which maximizes at about 40\,--\,50 months after the sunspot-cycle maximum. For the odd cycles the maximum is earlier some 10\,--\,20 months after the sunspot cycle maximum. These maxima are related to high-speed streams and coronal mass ejections, respectively, during the declining phase of the solar cycle \citep{Gosling_1977, Simon_1986, Simon_1989, Cliver_1996, Echer_2004}. Note also that the same phenomenon is seen also in the distributions of Dst-, Dxt-, and Dcx-indices (see Figures \ref{fig:Dst_histogram}, \ref{fig:Dxt_histogram}, and \ref{fig:Dcx_histogram}) but as a fainter feature. These phenomena affect the overall cross-correlation between the solar indices and geomagnetic disturbance indices such that the delay in the correlation is usually at least one or two years.

\section{IMF B and SW Velocity During Solar Cycles 20\,--\,24}

\begin{figure}
	\centering
	\includegraphics[width=0.9\textwidth]{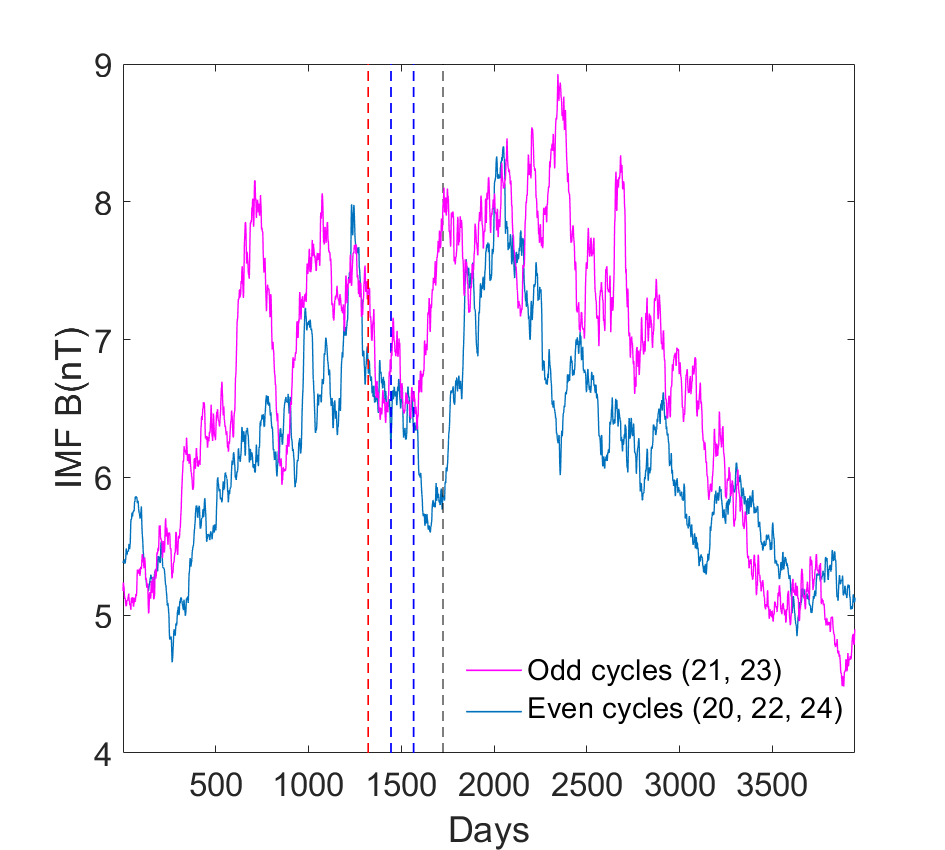}
		\caption{(Blue) The average absolute value of the smoothed magnetic-field intensity for the even Solar Cycles 20, 22, and 24. (Purple) Same but for the odd Solar Cycles 21 and 23.}
		\label{fig:IMF_B}
\end{figure}

Figure \ref{fig:IMF_B} shows the smoothed interplanetary magnetic-field magnitude average $\left|B\right|$-component from the daily OMNI2 data for the even Solar Cycles 20, 22, and 24, and for the odd Solar Cycles 21 and 23. The values were resampled similarly to the solar and geomagnetic data discussed earlier. The vertical lines are also plotted similarly to earlier figures. There is a decline in IMF absolute value of magnetic-field intensity \textit{B}-component for even cycles, and as deepest between 1567\,--\,1725 days from the start of the resampled average cycle. This resembles the decline in Dst-related indices in Figure \ref{fig:Dcx_Dxt_Dst_averages}.
For the even cycles the decline is between the lines at 1322\,--\,1567, with a little spike after 1445 days. This behavior is also similar to the Dst-, Dxt-, and Dcx-curves for odd cycles in Figure \ref{fig:Dcx_Dxt_Dst_averages}. The T-test shows that the gap in \textit{B} for the even cycles between 1445\,--\,1725 days has significantly different (smaller) mean value (6.42\,nT) at least at level 99\,\% with p\,$<\,$10$^{-6}$ than intervals nine months earlier (7.15\,nT) or nine months later (7.23\,nT). These values are calculated from the daily raw data, not from smoothed data. For the odd cycles the significance of the mean \textit{B} between 1322\,--\,1567 (6.79\,nT) is at the 99\,\% level with at least p\,=\,$4.1\times10^{-4}$ compared to eight months before (7.41\,nT) and after (7.43\,nT) the period. While the OMNI2-data hve been measured near Earth in geocentric or $L_{1}$ (Lagrange point) orbits, it is not surprising that the changes in solar activity and later in IMF soon appear as geomagnetic responses at the Earth.

\begin{figure}
	\centering
	\includegraphics[width=0.9\textwidth]{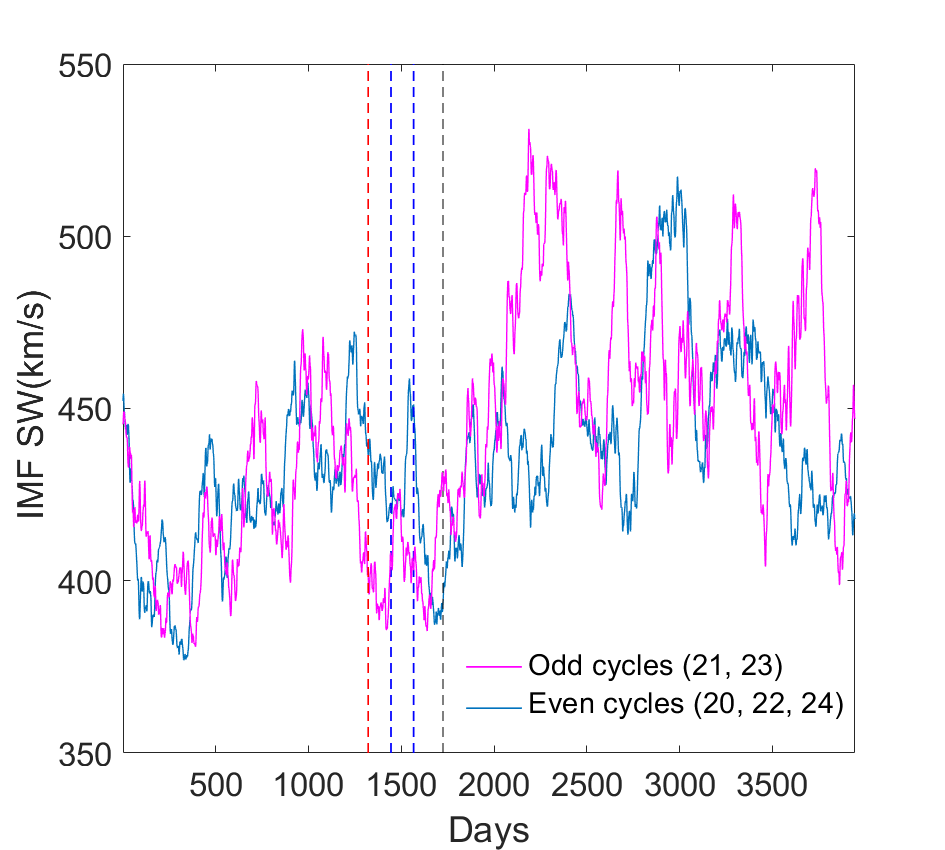}
		\caption{(Blue) The average smoothed solar-wind velocity for the even Solar Cycles 20, 22, and 24. (Purple) Same but for the odd Solar Cycles 21 and 23}
		\label{fig:IMF_SW}
\end{figure}

Figure \ref{fig:IMF_SW} shows the solar-wind average velocity during the even and odd Solar Cycles 20\,--\,24. Surprisingly, there is now a decline for the odd cycles between the whole interval lines at 1322\,--\,1725 (404.3 km$\texttt{s}^{-1}$), but for even cycles it is as deepest at the end of the previously used intervals, i.e. between 1567\,--\,1725 (403.6 km$\texttt{s}^{-1}$). Note also that the solar-wind velocity reaches a maximum similarly to the Ap-index in the descending phase of the cycles, and earlier for the odd cycles than for the even cycles. As mentioned earlier these maxima are related to coronal mass ejections and high-speed streams in the solar wind. Significance for different mean of the even cycles in the interval 1567\,--\,1725 compared to five months before the interval (445.4 km$\texttt{s}^{-1}$) is clear with p\,$<\,10^{-13}$, and it is still significant at the 99\,\% level with p\,=\,$1.7\times10^{-5}$ to five months after the interval (430.0 km$\texttt{s}^{-1}$). For the odd cycles the significance nine months before (435.2 km$\texttt{s}^{-1}$) and nine months after (436.7 km$\texttt{s}^{-1}$) the interval 1322\,--\,1725 days is very clear with p\,$<\,10^{-11}$ for both periods. Note that the solar wind behaves differently than the magnetic-field intensity such that the average level of the velocity is much higher, with occasional high peaks during the descending phase of the cycle.

\section{The Group Area, IMF $BV^{2}$, and Ap-Sequence for Solar Cycles 20-24}

\begin{figure}
	\centering
	\includegraphics[width=0.8\textwidth]{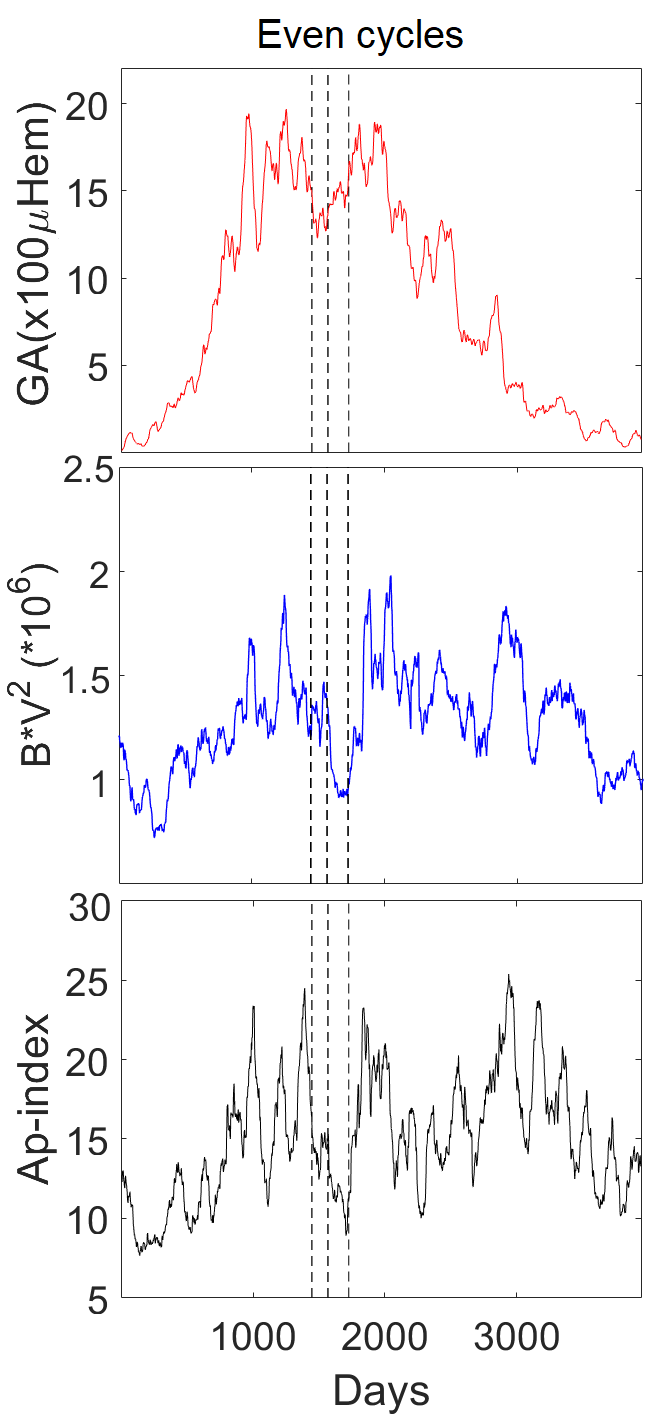}
		\caption{Top panel: The average sunspot-group area for the even Solar Cycles 20\,--\,24. Middle panel: The average IMF $BV^{2}$-component for the same cycles. Bottom panel: The average Ap-index for the same cycles.}
		\label{fig:Sun_IMF_Ap_even}
\end{figure}

Before concluding, we draw together the whole sequence of the group areas, a function IMF $BV^{2}$ \citep{Ahluwalia_2000}, and Ap-index during the intervals 1445\,--\,1725 and 1322\,--\,1567 for the even and odd Solar Cycles 20\,--\,24, respectively.
Figure \ref{fig:Sun_IMF_Ap_even} depicts the sequence of the GG phenomenon from the Sun through IMF to the Earth for the even cycles. Here all the curves are calculated only for the even Solar Cycles 20, 22, and 24. Note that there are two drops in the GG interval for the groups areas: first between 1445\,--\,1567 and second between 1567\,--\,1725 days. The first drop has only a small response in $BV^{2}$ and Ap-index, but the second one has a strikingly similar decline in both $BV^{2}$ and geomagnetic activity. Both $BV^{2}$ and Ap-index reach the smallest values at the end of this interval (except of course at the beginning and end of the cycle). Note, however, that the Ap minimum is sharper than the IMF $BV^{2}$ minimum.

\begin{figure}
	\centering
	\includegraphics[width=0.8\textwidth]{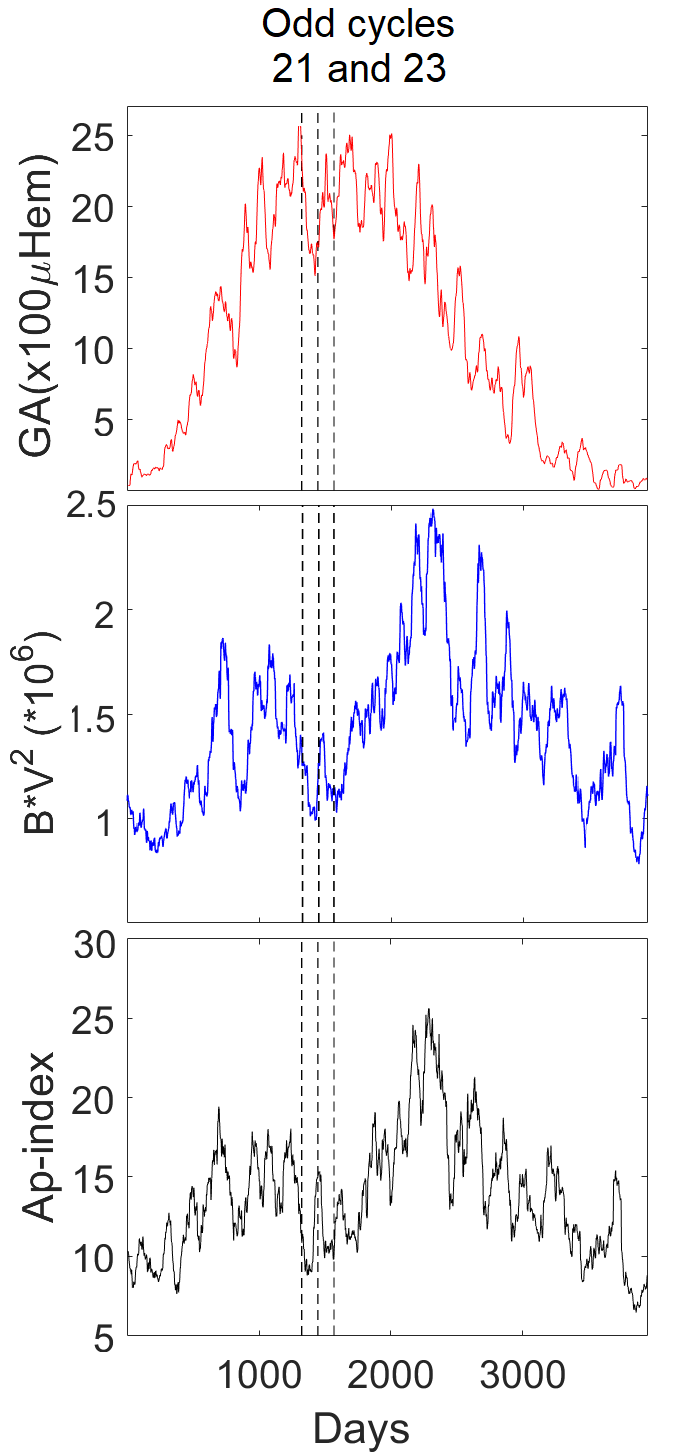}
		\caption{Top panel: The average sunspot-group area for the odd Solar Cycles 21 and 23. Middle panel: The average IMF $BV^{2}$-component for the same cycles. Bottom panel: The average Ap-index for the same cycles.}
		\label{fig:Sun_IMF_Ap_odd}
\end{figure}

Figure \ref{fig:Sun_IMF_Ap_odd} depicts the sequence of the GG phenomenon for the odd Solar Cycles 21 and 23. For these two cycles the decline in areas between 1322\,--\,1445 is larger than for the average area for odd cycles between 13\,--\,23 (see Figure \ref{fig:Sunspot_area}b). This is due to the fact, that the Solar Cycles 21 and 23 are more double-peaked than the earlier odd cycles. There are also a two-fold drop between 1322-1567 days after the cycle start for IMF $BV^{2}$ and Ap-index: first drop between 1322\,--\,1445 days and second drop between 1445\,--\,1567 days. Here the first drop is larger than the second drop for both $BV^{2}$ and Ap-index. Note that the decline in the function $BV^{2}$ continues still after the line at 1567, and this maybe leads to a third drop in the Ap-index.

\section{Conclusions}

We have shown that the time series of sunspot-group areas has a gap, the so-called Gnevyshev gap (GG), between ascending and descending phases of the solar cycle and especially so for the even cycles. For the odd cycles this gap is less obvious, and is only a short but sudden decline after the maximum of the cycle. We resampled the cycles to have same length of 3945 days (about 10.8 years), and we showed that the decline is between 1445\,--\,1567 days after the start of the cycle for the even cycles, and extending sometimes until 1725 days from the start of the cycle. For the odd cycles the gap is a little earlier, 1332\,--\,1445 days after the start of the cycles with no extension. We also find an approximately 150-day periodicity from the time series of the sunspot-area data. We believe that this is the same period that has been reported in several articles, especially in solar-flare events \citep{Rieger_1984, Lou_2000, Richardson_2005}. Furthermore, we showed that the smaller sunspot groups do not contribute significantly to this double-peak pattern in the solar cycle.

We analyzed geomagnetic disturbances for Solar Cycles 17\,--\,24 using Dst-index, related Dxt- and Dcx-indices, and Ap-index. In all of these time series there is a decline at the time or somewhat after the GG in the solar indices. These declines are 46\,\% smaller for Dst- and Dxt-indices during 1445\,--\,1725 than in the surrounding intervals for the even cycles, and are significant at 99\,\% level for both even and odd cycles of the Dst-index and for Dxt-, Dcx- and Ap-indices for even cycles. For odd cycles of the Dxt-index the significance is 95\,\%, but insignificant for odd cycles of Dcx- and Ap-indices. We have plotted histograms for the Dst-, Dxt-, and Dcx-indices in two categories: Index$<$\textrm{-}50\,nT and \textrm{-}50\,nT$\leq$Index$<$\textrm{-}15\,nT, and for Ap-index in categories: Ap$>$45\,nT and 15\,nT$<$Ap$\leq$45\,nT. The histograms show much fewer counts for the even cycles between 1445\,--\,1725 days (nine-month interval) from the start of the average cycles for all indices and in both categories. For odd cycles we discovered significantly fewer counts in both categories only for Dst-index, and slightly fewer counts in the ``stronger" disturbance categories for other indices. Note that Dst-index exists only for two odd cycles, 21 and 23, which partly affects this behavior.

The analyses of OMNI2 data for 1964\,--\,2020 (SC20\,--\,SC24) showed that these gaps during the maxima of thes solar cycles also exist in the near-Earth solar-wind velocity, IMF magnetic field intensity [\textit{B}], and the combined function $BV^{2}$. It is, however, not surprising that the GG phenomenon exists in the solar-originated IMF and solar-wind data. The significance of the decline during the GG interval is high for both even cycles and odd cycles in $\left|B\right|$ and in $BV^{2}$.

Interestingly, the solar-wind velocity behaves in a different way. There is a fall in the values simultaneously with the GG in the Sun, but the depression lasts longer than in IMF \textit{B} values and in geomagnetic indices for the odd cycles (between 1322\,--\,1725 days), and shorter for the even cycles (only between 1567\,--\,1725 days). If we divide the average cycle of SW in two halves, the first half has mean velocity 434.0 km$\texttt{s}^{-1}$ and 419.7 km$\texttt{s}^{-1}$ for even and odd cycles and the second half 465.4 km$\texttt{s}^{-1}$ and 467.5 km$\texttt{s}^{-1}$ for even and odd cycles, respectively. Especially for odd cycles, this difference is much larger than the GG-related decline in the solar-wind velocity. As we have discussed earlier, the fast velocities during descending phase of the cycle are due to the coronal mass ejections and high-speed streams in the solar wind.

\begin{acknowledgments}
The sunspot-group area data are downloaded from www2.\newline mps.mpg.de/projects/sun-climate/data.html. The dates of cycle minima were obtained from from the National Geophysical Data Center, Boulder, Colorado, USA (ftp.ngdc.noaa.gov). The Dst-index is published at wdc.kugi.kyoto-u.ac.jp\newline /dst\_final/index.html. Dcx- and Dxt-indices are retrieved from dcx.oulu.fi. \newline Definitive values of Ap are provided by the GeoForschungs Zentrum (GFZ) Potsdam. The OMNI2 data are downloaded from spdf.gsfc.nasa.gov/pub/data/omni/\newline low\_res\_omni/.
\end{acknowledgments}

\flushleft
\textbf{Disclosure of Potential Conflicts of Interest} \newline
\footnotesize {The author declares that there are no conflicts of interest.}

%
%
\bibliographystyle{spr-mp-sola}
\bibliography{references_JT_SolPhys}  
%
%
%
%


\end{document}